\documentclass[12pt]{article}
\usepackage{epsfig}
\newcommand{\tr}{{\mathrm{tr}}}

\newcommand{\nf}{{n_\mathbf{f}}}
\newcommand{\nd}{{n_\mathbf{d}}}
\newcommand{\ns}{{n_\mathbf{s}}}
\newcommand{\na}{{n_\mathbf{a}}}
\newcommand{\nv}{{n_\mathbf{v}}}
\newcommand{\nt}{{n_\mathbf{t}}}
\newcommand{\nr}{{n_\mathbf{R}}}

\newcommand{\be}{\begin{equation}}
\newcommand{\ee}{\end{equation}}

\expandafter\ifx\csname mathbbm\endcsname\relax

\else

\fi
\begin{document}
\begin{titlepage}
\begin{flushleft}
       \hfill                      {\tt hep-th/9703098}\\
       \hfill                      UUITP-7/97\\
       \hfill                       March 1997\\
\end{flushleft}
\vspace*{3mm}
\begin{center}
{\LARGE Notes on Supersymmetric Gauge Theories }\\
{\LARGE in Five and Six Dimensions}\\
\vspace*{12mm}
{\large Ulf H. Danielsson\footnote{E-mail: ulf@teorfys.uu.se},
 Gabriele Ferretti\footnote{E-mail: ferretti@teorfys.uu.se}, \\
\vspace*{3mm}
 Jussi Kalkkinen\footnote{E-mail: jussi@teorfys.uu.se}, and P\"{a}r
Stjernberg\footnote{E-mail: paer@teorfys.uu.se}\\
\vspace{5mm}
{\em Institutionen f\"{o}r teoretisk fysik \\
Box 803\\
S-751 08  Uppsala \\
Sweden \/}\\}
\vspace*{10mm}
\end{center}

\begin{abstract}
We investigate consistency conditions for supersymmetric gauge theories
in higher dimensions. First, we give a survey of Seiberg's necessary
conditions for the existence of such theories with simple groups in five
and six dimensions. We then make some comments on how theories in
different dimensions are related. In particular, we discuss how the
Landau pole can be avoided in theories
that are not asymptotically free in four dimensions, and the mixing of
tensor and vector multiplets in dimensional reduction from six dimensions.
\end{abstract}
\end{titlepage}

\section{Introduction}

One of the most intriguing spin-offs of the recent wave of activity
on non perturbative supersymmetric field and string theories has been the
discovery of interacting supersymmetric gauge theories in five and six
spacetime dimensions\footnote{The theories we discuss in this paper are
those with eight supercharges, corresponding to $N=(1,0)$,  $N=1$ and
$N=2$ supersymmetries in six, five and four spacetime dimensions
respectively.} [1-8].

These theories can be constructed from compactification of string
theory
(including its non perturbative forms of M and F-theory) on an
appropriate manifold [9-19]. As the manifold develops a
strong coupling singularity, the set of fields that are not localized on
the singularity, decouples from the physics,
leaving a theory in flat space time
with, possibly, some extra global symmetry \cite{W4}. The remaining fields
provide the
gauge and matter content necessary to describe the theory, with the
possible addition of tensor multiplets in six dimensions.

Seiberg has found \cite{Se1,Se2} some simple \emph{necessary}
conditions that must be
satisfied by the field theory in order to have such a non trivial fixed
point.
Seiberg's arguments are purely field theoretical and what is remarkable is
that
they have a counterpart in the geometry of the compactification, making it
quite plausible that the field theoretical requirements are also
sufficient.

In this paper we continue this line of analysis by considering two
separate issues:

First, we give an explicit survey of all candidate six dimensional
theories with simple gauge groups. This is a straightforward exercise in
the calculation of anomalies [20-27] and the situation can
be summarized as
follows. For those groups lacking an independent fourth order Casimir
\cite{O}
($SU(2)$, $SU(3)$ and all exceptional groups) the situation is
qualitatively similar to the one in five and four dimensions, i.e., there
is an ``upper bound'' on the amount of matter that is allowed in the
theory. In particular, pure gauge theories based on these groups are
always possible \footnote{See however Note Added.}. 
For all other gauge groups, there must be exactly the
``right amount'' of matter in order to satisfy the consistency condition.
For example, the only group in this category for which a
pure gauge theory is allowed is $Spin(8)$. As has already been noticed in
comparing
five to four dimensions \cite{IMS}, the conditions in six dimensions are
neither
stronger nor weaker than in the other cases. 

Second, we study some basic phenomena of ``dimensional crossover'' as we
compactify these theories in five and four dimensions. (These arguments
have a potentially wider range of applicability than the setting of this
paper.) Given the ``mismatch'' between the acceptable matter contents in
various dimensions, the issue arises on how these theories are related by
compactification on a circle. For instance, it is well known that an
$SU(2)$ theory can have up to $\nf=4$ matter hypermultiplets in the
fundamental representation in four dimensions, whereas Seiberg \cite{Se1}
has
shown that one can have up to $\nf=7$ in five dimensions\footnote{$\nf=8$
is a borderline case that does not have a strong coupling limit but must
also be taken into account under certain
circumstances.}. This means that,
by adding an extra compact dimension, (no matter how small), to the four
dimensional theory, it is possible to avoid the Landau pole for a
certain range of coupling constants. On the contrary, for an $SU(3)$ gauge
theory with one hypermultiplet in the symmetric representation
$\mathbf{s}=\mathbf{6}$ an extra
dimension would be disastrous, as the theory is asymptotically free in
four dimensions but ill-defined in five \cite{IMS}.

Going from five to six dimensions is even more challenging because of the
further restrictions on the prepotential and the existence of a new
multiplet in six dimensions: the tensor multiplet. We already know that it
is possible for a tensor multiplet to turn into a gauge field in the
Cartan subalgebra of a particular gauge group (other then $U(1)$!) upon
compactification \cite{GMS}.
The natural question that arises here is whether it is possible for this
field
to mix with the dimensional reduction of other gauge fields already
present in six dimensions yielding an enlarged gauge symmetry in five
dimension\footnote{This question was also raised in \cite{IMS}.}. We
present evidence against this phenomenon although this might
require further studies. Some other technical difficulties in the
dimensional reduction of tensor multiplets were discussed in \cite{PS} in
relation to effective actions for $p$-branes \cite{CVNW}.

The paper is organized as follows. In Section two we review some of the
results on the five dimensional theories \cite{IMS} for further reference;
in Section
three we discuss the consistency requirements for six dimensional theories
and finally, in Section four we discuss various aspects of dimensional
crossover.

\section{Review of five dimensional theories}

In this section we briefly review some of the results for five dimensional
supersymmetric gauge theories with eight supercharges and simple gauge
group $G$. This analysis was started in \cite{Se1} and extended in 
\cite{MS, GMS, IMS}.
There are two possible multiplets: the vector multiplet, whose real
scalar component we denote by $\phi^a$, $a=1, \cdots, \mathrm{dim}\; G$,
and a set of hypermultiplets transforming under a generic representation
$\mathbf{R_1}\oplus \mathbf{R_2}\cdots$ of the gauge group. Throughout
the paper, we denote the number of flavors in a certain
representation $\mathbf{R}$ by $\nr$ ; if $\mathbf{R}$ is pseudoreal and
no
global anomaly is
present, one is allowed to couple half a hypermultiplet to the gauge
field,
i.e., $\nr$ can be a half-integer.

The Coulomb branch is parametrized by the scalars $\phi^i$ in the vector
multiplet belonging to the Cartan subalgebra of $G$, $i=1, \cdots,
\mathrm{rank}\;G$ and it is topologically a wedge given by modding out the
Cartan subalgebra by the action of the Weyl group. 
In \cite{IMS} it is shown that the necessary
condition for the existence of a non trivial UV fixed point is that, in
the limit $g_0=\infty$, the quantum prepotential
\be
    \mathcal{F} = \frac{1}{2g_0^2}\delta_{ij}\phi^i\phi^j +
    \frac{c_\mathrm{class.}}{6} d_{ijk} \phi^i\phi^j\phi^k +
    \frac{1}{12}\left(\sum_\alpha|\alpha_i \phi^i|^3 -
    \sum_\mathbf{R} \nr \sum_{w\in \mathbf{R}}|w_i\phi^i +
     m_\mathbf{R}|^3 \right) \label{prepot}
\ee
must be a convex function throughout the Coulomb branch
\footnote{Excluding the case of $\mathcal{F}$ identically zero
in this limit; i.e. excluding the presence of
adjoint matter.}.
($\alpha$ denotes all the roots of $G$, $w$ all the weights of
$\mathbf{R}$ and $d_{ijk}$ is the third rank symmetric invariant tensor
that exists only for the groups $SU(N)$ with $N\geq 3$;
$c_\mathrm{class.}$ is quantized by considering the
global anomaly associated with $\pi_5(SU(N))=\mathbf{Z}$.)
The analysis of \cite{IMS} yields the following results:

$\bullet$ For $SU(N)$ only the fundamental representation $\mathbf{f}$ and
the
antisymmetric two tensor $\mathbf{a}$ are allowed and the quantization
condition is
\be
c_\mathrm{class.} + (\nf + N \na)/2 \in\mathbf{Z}.
\ee
For $N>8$ only $\na=0$ and $\nf + 2|c_\mathrm{class.}|\leq
2N$ are allowed. For $5\leq N \leq 8$, $\na=1$ and
$\nf+2|c_\mathrm{class.}|\leq 8-N$ is also allowed and, for $N=4$,
$\na\equiv n_\mathbf{6}=2$ and $\nf\equiv
n_\mathbf{4}=c_\mathrm{class.}=0$ is also allowed.
In the degenerate cases of $SU(2)$ and $SU(3)$, where the antisymmetric
representation is either trivial or conjugate to the fundamental, one
finds
$\nf \equiv n_\mathbf{2}\leq 7$ and $\nf \equiv n_\mathbf{3}\leq 6$
respectively.

$\bullet$ For $Sp(N)$ ($N\geq 2$ being $Sp(1) = SU(2)$), one again finds
that the only possible representations are $\mathbf{f}$ and $\mathbf{a}$
with the
requirements
$\na=0$, $\nf\leq 2N+4$ or $\na=1$, $\nf\leq 7$.
The second case reduces the system to a direct product of $SU(2)$ groups
as already familiar from four dimensions [32-35], where a
non-renormalization theorem on the Higgs branch can be used.

$\bullet$ For $Spin(N)$ ($N\geq 7$ being $Spin(6) = SU(4)$ etc...) the
only
representations allowed are the vector $\mathbf{v}$
and the spinor $\mathbf{d}$. One always has $\nv \leq N-4$ and,
for $N=7,8,9,10,11,12$, one can also have $\nd \leq 4,4,2,2,1,1$ in the
same
order.

$\bullet$ The exceptional groups have only one representation
``$\mathbf{f}$''
smaller than the adjoint, except for $E_8$ that has none. The bounds are
$\nf \leq 4,3,4,3$ for $G_2$, $F_4$, $E_6$ and $E_7$. In the case
of $E_7$ one can have an odd number of half-hypermultiplets. $E_8$ cannot
be
coupled to matter in this way.

\section{Consistency conditions in six dimensions}

In six dimensions, Seiberg's necessary condition \cite{Se2} is that, after
gravity has
decoupled, the gauge anomaly can be canceled by the introduction of at
least one tensor multiplet. The contribution to the anomaly eight-form
from
the gauge multiplet and the hypermultiplets is
\be
I_8 = \tr_\mathbf{Adj.} F^4 - \sum_\mathbf{R} \nr \tr_\mathbf{R} F^4
\equiv
\alpha \tr_\mathbf{f} F^4 + c(\tr_\mathbf{f} F^2)^2.
\ee
To be able to cancel this part of the anomaly without introducing gravity
and with at least one tensor multiplet we  need $\alpha=0$ and
$c > 0$\footnote{The case $c=0$ need also be considered in certain
cases.}.
This requirement is very easy to investigate.

The simple groups can be divided into two classes. The first class
consists of those groups that do not have an independent fourth order
Casimir \cite{O}.
They are $SU(2),SU(3)$ and all the exceptional groups $G_2$, $F_4$,
$E_6$, $E_7$ and $E_8$. For these particular groups
$\tr_\mathbf{Adj.}F^4$ and $\tr_\mathbf{R}F^4$
can always be expressed
in terms of $\tr_\mathbf{f}(F^2)^2$ and $\alpha=0$.
The anomaly condition $c>0$
becomes an upper bound on the number of matter hypermultiplets, and,
hence,
closer in spirit to the results in four and five dimensions.
All other groups possess a fourth order Casimir, and the requirement
$\alpha=0$
implies that one must have just the right amount of matter.

We first consider groups in the first category and investigate the
possible matter content. We need only consider
representations whose dimension is smaller than the dimension of the
adjoint. This restricts the possibilities to the fundamental
representations, except for $SU(3)$, where we also may have the
symmetric $\mathbf{s}=\mathbf{6}$. The groups $SU(2)$ and $E_7$ have
pseudoreal
fundamental
representations which allows for a half-integer number of
hypermultiplets. 

Using the tables of \cite{E} we can calculate the anomaly polynomial
and find that $c$ is proportional to

\begin{eqnarray}
16 -  n_{\bf 2} & \mathrm{ for } &SU(2) \nonumber \\
18 - n_{\bf 3} - 17n_ {\bf 6} &  \mathrm{ for } &SU(3) \nonumber \\
10 - n_ {\bf 7} &  \mathrm{ for }  &G_2 \nonumber \\
5 - n_ {\bf 26}  & \mathrm{ for } &F_4 \nonumber \\
6 - n_{\bf 27} & \mathrm{ for } & E_6 \nonumber \\
4 - n_{\bf 56}&  \mathrm{ for } &E_7,
\end{eqnarray}
from which one immediately reads off the required bounds. Note that the
$\mathbf{6}$ of $SU(3)$, that is allowed by asymptotic freedom in four
dimensions and forbidden in five, has reappeared in six dimensions, where
one can have $n_{\bf 3}=0$ and $n_ {\bf 6}=1$\footnote{This particular example,
and a few others based on the groups $SU(2)$, $SU(3)$ and $G_2$, have been
shown to be affected by a global anomaly, see Note Added.}.

Let us then turn to the second class of simple groups for which there
is an independent fourth order Casimir. In this case we have to solve
both equations $\alpha =0$ and $c>0$. The result is no longer a bound
for the maximum number of matter multiplets, but the matter content
must exactly compensate the contributions of the gauge sector. There
is no global anomaly as $\pi_6(G)$ is trivial in all these cases.

$\bullet$ For $SU(N)$, $N > 3$, we need only consider the case of $\nf$
multiplets in the fundamental representation, $\ns$ in
the second rank symmetric, $\na$
in the second rank antisymmetric and $\nt$
in the third rank antisymmetric.
For $N\geq 8$, we can have the three combinations:
\be
   (\nf,\na,\ns,\nt)=(2N, 0,0,0), \; (N+8,1,0,0), \; (N-8,0,1,0).
\ee
For $N=7$ we find only:
\be
   (\nf,\na,\ns,\nt)\equiv(n_\mathbf{7}, n_\mathbf{21},
    n_\mathbf{28}, n_\mathbf{35})=(14, 0,0,0), \; (15,1,0,0).
\ee
For $N=6$ one can finally have\footnote{This is the only case where
$\mathbf{t}$ is allowed. Also, in this case, the anomaly cancels
completely for
$(n_\mathbf{6}, n_\mathbf{15},  n_\mathbf{21}, n_\mathbf{20}) = (0,1,1,0),
(16,2,0,0), (17,1,0,1/2), (18,0,0,1)$ and $(1,0,1,1/2)$.}:
\be
   (\nf,\na,\ns,\nt)\equiv(n_\mathbf{6}, n_\mathbf{15},
    n_\mathbf{21}, n_\mathbf{20})=(12, 0,0,0), \; (14,1,0,0), \;
    (15,0,0,\frac{1}{2}).
\ee
For $SU(4)$ and $SU(5)$ the third rank representation $\mathbf{t}$ is
conjugate to the fundamental $\mathbf{f}$ and to the antisymmetric
$\mathbf{a}$ respectively, and we also find $\ns =0$; hence
\begin{eqnarray}
   (\nf,\na)\equiv(n_\mathbf{4}, n_\mathbf{6}) = (8,0), \; (12,1)
    & \mathrm{ for } &SU(4) \nonumber \\
   (\nf,\na)\equiv(n_\mathbf{5}, n_\mathbf{10}) = (10,0), \; (13,1)
    &  \mathrm{ for } &SU(5).
\end{eqnarray}

$\bullet$ Of the $Spin(N)$, $N \geq 7$, representations we consider the
vector
representation $\mathbf{v}$, the spin representation $\mathbf{d}$, the
symmetric second rank representation $\mathbf{s}$ and the antisymmetric
third rank representation $\mathbf{t}$. It turns out that there is a
solution only in the absence of the  ${\bf s}$ and  ${\bf t}$
representations. For $Spin(7)$ we get three possibilities $(\nv,
\nd)\equiv(n_\mathbf{7}, n_\mathbf{8})=(0,2),(1,4)$ and $(2,6)$. For
$N=8,\ldots,12$ there are four possibilities:
\begin{eqnarray}
\nv &=& N-8, \quad \nd = 0 \nonumber \\
\nv &=& N-7, \quad \nd = 2^{5-[(N+1)/2]} \nonumber \\
\nv &=& N-6, \quad \nd = 2^{6-[(N+1)/2]} \nonumber \\
\nv &=& N-5, \quad \nd = 3 \cdot 2^{5-[(N+1)/2]},
\end{eqnarray}
where the brackets $[~]$ denote the integer part. Notice in particular
that for $Spin(8)$ we find a solution for $c>0,\alpha=0$ without
introducing any matter. For $Spin(13)$ we get $(\nv, \nd)
\equiv(n_\mathbf{13}, n_\mathbf{64})=(5,0)$ and $(7,1/2)$. For $N\geq 14$
the spinor representations $\mathbf{d}$ are no longer allowed, and the
solution is
$\nv = N-8$.

$\bullet$ For $Sp(N)$, $N \geq 2$ we consider matter in the fundamental
$\mathbf{f}$ and
both in the second and in the third rank antisymmetric representations
$\mathbf{a}$ and $\mathbf{t}$. There is a solution only for $2N + 8$
multiplets in the fundamental representation and none in the
antisymmetric representations. The anomaly cancels completely for
$(\nf,\na,\nt)=(16,1,0)$ for any $N \geq 2$,
and $(\nf,\na,\nt)\equiv(n_\mathbf{6},n_\mathbf{14},n_\mathbf{14'})
=(17,0,1/2)$ for $N=3$.

\section{Dimensional crossovers}

\subsection{Going from four to five dimensions}

As we have seen, the consistency requirements for gauge theories in
different dimensions show a very complex behavior. A theory that is
admissible in a
lower dimension might be ill-defined in a higher dimension and vice versa.
It is
interesting to study the behavior of such theories as we change
the effective number
of  dimensions. Let us consider compactifying a five dimensional theory on
a circle;
The dimensionless quantity that determines the effective
number
of dimensions is $\phi R$, where $R$ is the radius of compactification
and $\phi$ the magnitude of the Higgs field on the Coulomb branch.

Let us study a couple of examples in detail. We begin with the case of
gauge group
$SU(2)$ with $\nf$ fundamental hypermultiplets. When $\nf <4$, we
assume
$1/R$ and $\phi$ are both $>> \Lambda$ so that we can ignore instanton
effects; The full solution in the IR is, of course, given in
Ref.~\cite{SW1,SW2},
and it is well defined for small $\phi R$ as well.

Using the results of \cite{N}, we find that the effective coupling is
given by summing up all the one loop contributions coming from the
Kaluza-Klein modes on the circle:
\be
\frac{1}{g^2_{eff}} =\frac{1}{g^2_{0}} +4\log (\sinh ^2  (2\pi R \phi ))
- \nf \log (\sinh ^2  (\pi R \phi )) .
\ee
At small $\phi R$ we recover
\be
\frac{1}{g^2_{eff}} =\frac{1}{g^2_{0}} +4\log(2\pi R \phi )^2 - \nf \log
 (\pi R \phi )^2   ,
\ee
i.e. an effectively four dimensional theory.
In this case the theory is asymptotically free for $\nf  <4$.
In the infrared, i.e. small $\phi R$, the
full solution guarantees that $1/g^2_{eff}$ remains positive.
For  large $\phi R$ we find
\be
\frac{1}{g^2_{eff}} =\frac{1}{g^2_{0}} +(8-\nf ) 2 \pi R\phi   ,
\ee
i.e. an effectively five dimensional theory. The dimensionful five
dimensional
coupling $g_5$ is related to the one above through $1/g^2_{eff} = 2 \pi
R/g^2_5$.
In this case a sensible UV-theory needs $\nf \leq 8$ in order not to hit
a Landau
pole. The combined behavior is illustrated in figure 1.

\begin{figure}
\begin{center}
\epsfig{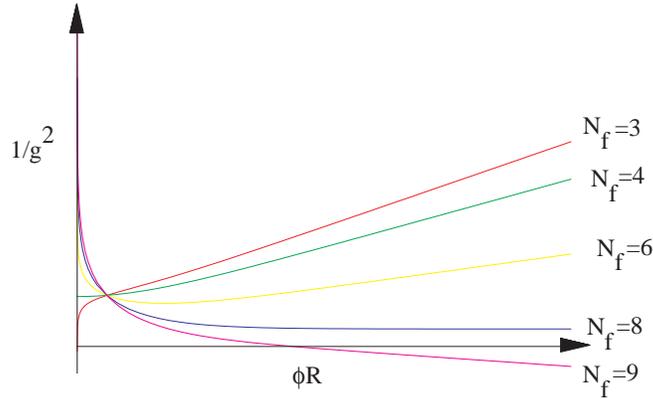}
\caption{The effective coupling for $SU(2)$ with $\nf$ fundamental
hypermultiplets on ${\bf R}^4 \times S^1$.}
\end{center}
\end{figure}

At a particular value of $\phi << 1/R$,  we have renormalized the value
of
the coupling to the same fixed value for all values of  $\nf$. If the
coupling is
small enough we find that the curves remain above zero for values $\nf
\leq 8$ and for this we would like to give the following physical
interpretation.

It is commonly believed that IR-free gauge theories in four dimensions,
like
$SU(2)$ with more then $4$ hypermultiplets in the fundamental
representation, are inconsistent
in the UV and will hit a Landau pole.
As long as the number of hypermultiplets are not greater than $8$
we see that these theories are in fact saved if there is a fifth dimension
that
opens up. In the UV the original four dimensional theory connects
smoothly
to the IR limit of a five dimensional theory. This, when we go ever
deeper
into the UV, has a UV-fixed point as described in \cite{Se1}.
If we measure the effective coupling to be $g$ at $\phi \sim \mu$ we
can estimate that an $R$ such that
\be
R> \frac{1}{\mu} e^{-\frac{1}{2(\nf-4)} \frac{1}{g^2}}
\ee
is needed to avoid a Landau pole.

To conclude, theories with $\nf <4$ are ``saved by instantons'' in the IR
and
theories with
$4 <\nf \leq 8$ are ``saved by the fifth dimension'' in the UV.

There are also examples of theories that are asymptotically free in four
dimensions but ill-defined
in five. Such an example is $SU(3)$ with a ${\bf 6}$. In this case, if we
consider fixing $R>0$, there is always a $\phi$ large enough that the
theory
hits a Landau pole along some direction on the Coulomb branch. The only
way to have $SU(3)$ with a ${\bf 6}$ is
therefore to have $R \equiv 0$.

\subsection{Going from five to six dimensions}

What happens to the six dimensional theories that we have studied when
dimensionally reduced to five dimensions? The tensor multiplet will
clearly give rise
to a vector multiplet in five dimensions and one might hope that it will
combine with the vector multiplet already present in six dimensions to
give interesting five dimensional physics. Let us investigate the
possibility for this to happen.

The question can be formulated in the following way; let $N_T$ be the
number of tensor multiplets and $G$ the gauge group already present in six
dimensions. After dimensional reduction to five dimensions, the Coulomb
branch becomes
$(N_T + \mathrm{rank}\; G)$-dimensional and the question is whether the
theory can be identified with an acceptable five dimensional theory based
on a simple group $\tilde G$, s.t., $\mathrm{rank}\; \tilde G = N_T
+ \mathrm{rank}\; G$.

Let us take $N_T=1$ and $G=SU(2)$ for simplicity, compactify on a circle
of radius $R$ and go far out in the Coulomb branch, where naive
dimensional reduction is allowed. In six dimensions we have, symbolically,
an interaction term
$\cal TAA$, where $\cal T$ and $\cal A$ are the tensor and vector
multiplets. This gives rise to an interaction that is locally of the type
${\cal A}_1{\cal A}_2{\cal A}_2$ in five dimensions, where ${\cal A}_1$
is the vector multiplet arising from the dimensionally reduced tensor
multiplet and ${\cal A}_2$ comes from the vector already present in six
dimensions. In addition, we could also have a term
${\cal A}_2{\cal A}_2{\cal A}_2$ that would be invisible in six
dimensions.
Cubic terms containing more than one $\mathcal{A}_1$ are forbidden on
dimensional grounds.

Therefore, if we consider the prepotential ${\cal F} = {\cal A}_1{\cal
A}_2{\cal A}_2 + b {\cal A}_2{\cal A}_2{\cal A}_2$ and compute the matrix
of coupling constants
\be
   \tau = \pmatrix{0 & \phi_2\cr \phi_2 & \phi_1 + 3b\phi_2 \cr},
\ee
we see that it has always a negative eigenvalue. We are therefore not
able to find an acceptable five dimensional theory after dimensional
reduction.

\section*{Note Added}

The correct treament of global anomalies has subsequently 
been given in \cite{BeVa}, (see also \cite{W3, EN}). The groups 
affected by the global anomaly are $SU(2)$, $SU(3)$ and $G_2$ and the 
allowed matter content is $n_\mathbf{2} = 4,\; 10$ for $SU(2)$, 
$n_\mathbf{3} = 0,\; 6,\; 12$ and $n_\mathbf{6} = 0$ for $SU(3)$, and 
finally $n_\mathbf{7} = 1,\; 4,\; 7$ for $G_2$.

\end{document}